# Comparative study of microscopy methods to assess fish intestinal microvilli

Ankit Butola[1, #], Luis E. Villegas-Hernández[1,2, #], Dhivya B. Thiyagarajan[3], Bartłomiej Zapotoczny[4] Roy A. Dalmo[3, $], and Balpreet Singh Ahluwalia[1,5 *]

*[1]Department of Physics and Technology, UiT The Arctic University of Norway, Tromsø, Norway*
*[2]Chip Nanoimaging AS, Tromsø, Norway*
*[3]Norwegian College of Fishery Science, Faculty of Biosciences, Fisheries and Economics UiT The Arctic University of Norway, Tromsø, Norway*
*[4]Department of Biophysical Microstructures, Institute of Nuclear Physics, Polish Academy of Sciences, Kraków, Poland*
*[5]Division of Obstetrics and Gynecology, Department of Clinical Science, Intervention and Technology, Karolinska Institute, Stockholm, Sweden*
*[#]Contributed equally*
*Corresponding authors: [$] roy.dalmo@uit.no [*]Balpreet.singh.ahluwalia@uit.no*

## Abstract:

The primary function of intestinal microvilli is to increase the surface area of the intestinal lining to maximize nutrient absorption. This is especially important as fish, like other animals, need to efficiently absorb proteins, carbohydrates, lipids, vitamins, and minerals from their digested food to support their growth and energy needs. Despite its importance to the fish health, the small size and dense footprint of microvilli hinders its investigation and necessitates the need of advanced microscopy methods for its visualization. Characterization of the microvilli using super-resolution microscopy provides insights into their structural organization, spatial distribution, and surface properties. Here, we present a comprehensive investigation of different optical, electron and force microscopy methods for analysis of fish microvilli. The super-resolution optical microscopy methods used are 3D structured illumination microscopy (SIM), stimulated emission depletion microscopy (STED), and fluorescence fluctuation based super-resolution microscopy (FF-SRM). We also visualized the intestinal microvilli in fish using diffraction-limited optical microscopy methods including confocal and total internal reflection fluorescence microscopy. Additionally, label-free microscopy methods, such as quantitative phase microscopy (QPM) and bright-field imaging, were also employed. To obtain ultra-high resolution, we used scanning electron microscopy (SEM), transmission electron microscopy (TEM) and atomic force microscopy (AFM). We demonstrate a systematic comparison of these microscopy techniques in resolving and quantifying microvilli features, ranging from 1-2 µm structural morphology to 10-100 nm surface details. Our results highlight the advantages, limitations, and complementary nature of each method in capturing microvilli characteristics across different scales. These techniques are used for super-resolution and high-resolution imaging of Atlantic salmon microvilli gastrointestinal tract. We believe that this methodological framework will serve as a valuable reference and provide the groundwork for future applications aimed to enhance fish health monitoring and other biological research in general.

## 1. Introduction:

The microvilli brush border is a dense physiological structure present in diverse organ tissues of vertebrates such as the placenta[1], the kidney[2], and the intestine[3]. This structure is critical for maximizing the efficiency of nutrient uptake, as well as gas and waste exchange by increasing the absorptive surface area at the interface between the nutrient environment and the tissues[3,4]. The traditional way of imaging the ultrastructure of microvilli ultrastructure is using scanning electron microscopy (SEM) and transmission electron microscopy (TEM)[5,6,7]. In the last few decades, several papers have been published to visualize nanometric structure of fish microvilli using TEM and SEM[6 8,9]. The primary advantage of these techniques lies in their ability to achieve super-resolution imaging at a scale of 1-20 nm that enables visualization of nanometric features of microvilli[7]. Due to this advantage, SEM and TEM have been used for various applications such as imaging gastrointestinal tract, gut microbiota, and among other[5,9].

The superior resolution of electron microscopes, however, comes with some limitations such as complexity in sample preparation and complexity of the instrumentation. This makes the imaging process labor intensive and time consuming. In addition, electron microscopes cannot capture the dynamics of cellular processes and lack the contextual spatial information that is essential to understand the biomolecular distribution of the biological samples. Recent advancements in optical microscopy offer novel opportunities to quantify nanometric structures of microvilli across a range of spatial scales[10-13]. The recent improvement in the resolution of optical microscopy helps bridging the gap between the electron microscopy and the conventional optical microscopy approaches to develop a better understanding of microvilli organization.

In this study, we present a systematic analysis of fish microvilli using electron microscopy methods, SEM, and TEM, atomic force microscopy (AFM), and different optical microscopy systems. In total seven different types of optical microscopy systems namely bright field microscopy, quantitative phase microscopy (QPM), on-chip total internal reflection fluorescence microscopy (TIRF), confocal microscopy, structured illumination microscopy (SIM), fluorescence fluctuation based super-resolution microscopy (FF-SRM), and stimulated emission depletion (STED) microscopy were used to compare the performance. This work aims to establish grounds for choosing optimal microscopy techniques for intestinal microvilli imaging fish intestinal tissues. Through this methodological approach, we aim to highlight the advantages, limitations, and complementary nature of these microscopy techniques in capturing microvilli characteristics at different spatial and structural scales. Our results emphasize the utility of combining optical, super-resolution, and electron microscopy techniques for a complete understanding of microvilli organization.

## 2. Overview of imaging methods

Advancement in imaging techniques improved our ability to observe and quantify various biological samples. Each imaging technique comes with its strengths and limitations while compared in terms of lateral resolution. Performance factors such as resolution, imaging throughput, specificity and ease of use play a significant role in determining the suitability of any imaging method for specific application. Here, we compared different methods used in this article in terms of their resolution, throughput imaging, specificity, quantitative imaging, speed, and ease of sample preparation as shown in Fig. 1. These parameters are chosen to highlight the strengths and weaknesses of each method for microvilli imaging. For example, the conventional bright-field imaging method is straightforward in throughput and sample preparation with high temporal resolution. However, it lacks the resolution, chemical specificity, and quantitative imaging capabilities required to visualize morphological changes in nanometric structures such as microvilli. On the other hand, TEM offers the best resolution among existing imaging systems but lacks chemical specificity and ease of sample preparation.

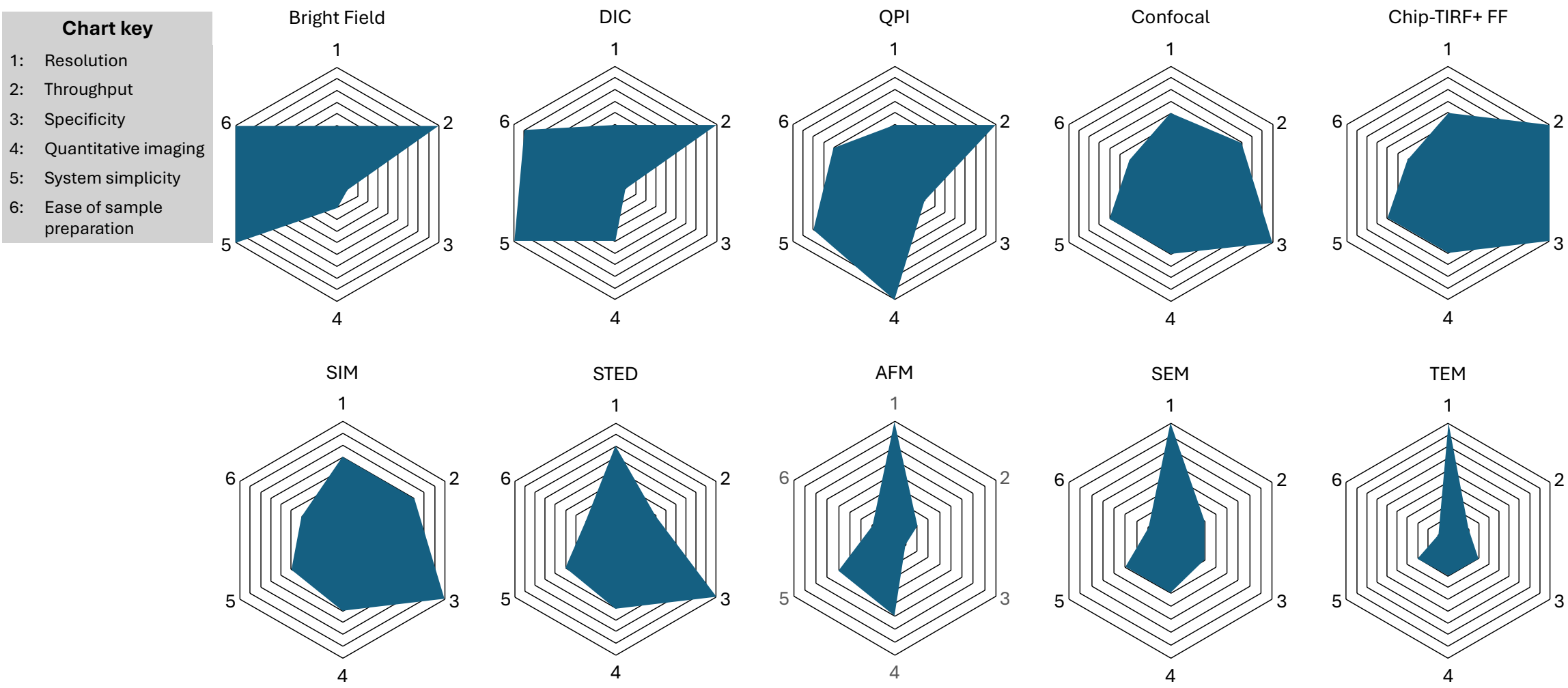

**Figure 1**: **Qualitative comparison of different optical and electron microscopy techniques using radar charts**. Six key parameters: resolution, throughput, chemical specificity, quantitative imaging, system simplicity, and ease of sample preparation are compared here. Each chart highlights the performance of a different technique, their strengths, and weaknesses. The radar chart has been prepared in the order of increasing resolution, which often comes with increasing complexity of sample preparation and reduction in the throughput.

### 2.1. Optical microscopy methods for microvilli visualization

The optical microscopy methods can broadly be categorized into label-free and label-based approaches. Label-free imaging involves visualizing biological specimens in their native state i.e., free from any external labeling agents. Label-free approach preserves the natural structure and dynamics of the sample while labelled-based approaches provide better resolution and chemical-specific information of the sample. These methods are explained and described in the next section.

#### 2.1.1. Label-free optical microscopy techniques:

A range of label-free techniques has been developed over the past few decades for example, bright-field, dark-field, phase contrast[14], differential interference contrast (DIC)[15], optical coherence tomography (OCT)[16], and quantitative phase microscopy imaging[17-20]. Label-free microscopy methods can be broadly categorized into two types based on their imaging capabilities: qualitative imaging and quantitative imaging (Fig. 2 (a)). In qualitative imaging, the light beams pass through the sample to generate the intensity image of the object. Techniques such as bright field, dark field, phase contrast, and differential interference contrast fall under this category[14,15]. These imaging techniques have been used for many biological applications such as live cell imaging, in-vitro fertilization, screening of blood cells screening, plant research, among others. However, due to the inherently semi-transparent and low-contrast nature of biological specimens, qualitative imaging methods suffer with limited contrast and structural details. This limitation arises because traditional intensity-based methods primarily rely on amplitude changes in transmitted or reflected light which are often subtle in biological cells and tissues[15].

To address these shortcomings, various quantitative label-free microscopy techniques such as OCT and QPM have been developed[20,21]. OCT and QPM utilize phase shift induced by light passing through the sample to extract their morphological and biophysical information[22]. To extract the phase shift, an additional reference beam interferes with the sample beam to produce an interferogram (Fig. 2 (b)). The interferometric image is processed through different computational algorithms to generate phase image that represent the optical thickness and refractive index map of the sample. The phase image of the sample can be useful to extract different quantitative parameters including dry mass, volume, and surface area with nanoscale sensitivity over an arbitrary period of time[18,19]. QPM methods also help accurately measure cell growth rates, which can be useful in addressing fundamental biological questions such as how cell growth is regulated and how cell size distribution is maintained[17,21]. The unique capabilities of QPM can also help address the long-standing debate on whether a cell's growth rate remains constant throughout its life cycle or scales proportionally with its mass. In recent years, different QPM techniques have been developed and used for several biomedical applications including screening of blood cells[23], ocular disease, breast cancer, in-vitro fertilization (IVF)[17], bacteria imaging, wound healing and among others[24,25].

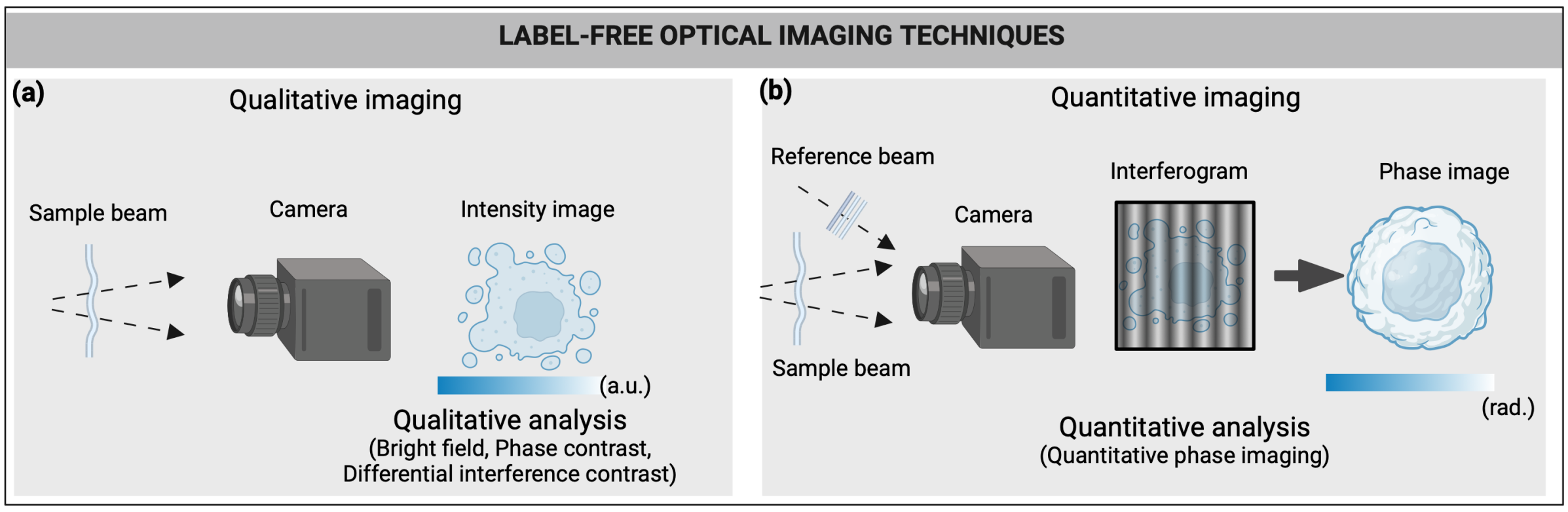

**Figure 2**: **Principle of label-free optical microscopy methods**: qualitative imaging and quantitative imaging. (a) In qualitative imaging, the sample is illuminated by the incident light, and the scattered field is recorded by the camera. Therefore, qualitative imaging provides intensity image of the sample. (b) In quantitative imaging, scattered light from the sample interferes with an additional beam known as the reference beam. The interreference encodes phase information i.e., refractive index and thickness maps which can be extracted by using computational algorithms. While qualitative imaging provided only intensity-based representation, quantitative imaging is useful to extract high-contrast images and morphology of the sample.

The limitations of QPM includes diffraction-limited resolution, reconstruction artefacts for thick and heterogenous samples, and a lack of chemical specificity[26]. The former restricts their ability to quantitatively visualize nanometric features, while the latter limits their applicability in pinpointing molecular changes. Recently, artificial intelligence (AI) has been used to address the lack of chemical specificity in QPM by virtual staining of phase images[27]. In these methods, a trained neural network is used to pinpoint different biomolecular distributions inside the cells. Although this is an interesting approach to identify and segment different intracellular organelles, careful examination is required to assess the accuracy of the resulting image. For example, determining the boundary between the nucleus and mitochondria is very crucial for calculating the accurate dry mass of the mitochondrial region in sperm cells[28]. From a computational perspective, it may be beneficial for the QPM community to adopt a generalized framework for training and testing of the networks to avoid artefacts, oversampling, and under-sampling in the target image. The experimental approach to tackle the challenge of molecular specificity in label-free imaging is to use Raman imaging, infrared imaging, multiphoton imaging, and multi-harmonic imaging[29-31]. Yet even these methods are limited by their resolution compared to the fluorescence imaging as described in the next section.

### 2.1.2. Fluorescence based optical imaging techniques:

The limitation of chemical specificity and resolution in label-free methods can be addressed using label-based optical imaging i.e., fluorescence microscopy. Fluorescence microscopy is a popular technique among biologists for visualizing specific structures within cells and tissues[32]. These methods can be used to target and visualize specific structures or molecules by employing fluorescence dyes and proteins. The versatility of fluorescence microscopy has made it a standard technique in many biological studies that require localization of cellular and subcellular components[32]. Over the last few decades, fluorescence microscopy has undergone significant advancements to address technical challenges such as suppressing the out of focus light, improving spatial resolution, axial sectioning, and imaging thick samples etc.[33,34]. For example, confocal microscopy is a type of fluorescence microscopy technique that uses a pinhole aperture to block out of focus light to provide better depth resolution and diffraction limited lateral resolution within the sample[35,36]. This unique capability of confocal microscopy makes it feasible for diverse applications, including imaging the intestinal valve and brush border in fish microvilli[37].

Another fluorescence microscopy method that generates images with excellent signal to noise is TIRF imaging[38]. TIRF enables selective excitation of fluorophores within a thin evanescent field to minimize the background noise and enhance signal-to-noise ratio[39,40]. This technique is remarkably effective for studying membrane associated process with minimal photobleaching. Recent advancements in TIRF imaging have shown high spatial bandwidth by using photonic chip-based geometries. Chip-TIRF is a comparatively new technique [41-44] where imaging high-contrast ultra-large field of view imaging is possible. This technique uses a waveguide platform to hold and illuminate the sample. In addition, it provides the flexibility to integrate the system with other imaging modalities where multimodal imaging of similar field of view can be possible. Using waveguide chip illumination, different super-resolution optical microscopy have been implemented such as photonic chip-based SIM, FF-SRM, and SMLM. The chip-TIRF has been extensively used for various biological studies including imaging liver cells, macrophages, tissues, and histopathological samples[18,41-44]. However, TIRF itself is insufficient for identifying individual microvilli due to the diffraction-limited nature of this imaging method. Since the feature size of microvilli ranges from 50-200 nm, they cannot be visualized using diffraction-limited imaging techniques.

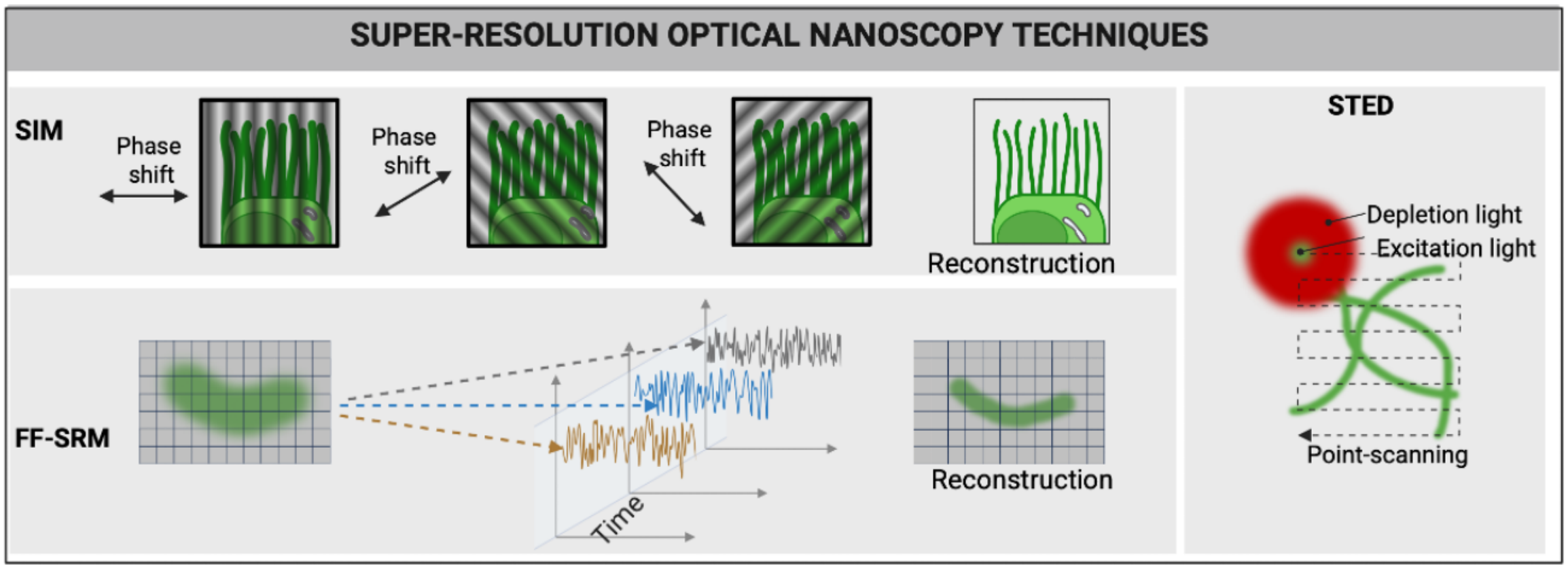

**Figure 3**: **Schematic diagram of working principle of different types of super-resolution techniques used in the present study for microvilli imaging**. 3D-SIM acquires 15 images to reconstruct a single super-resolved image, i.e. interference fringes in three angles and 5 phase steps per angle. FF-SRM uses a few hundred images to harness fluctuations in the fluorescence intensity to generate a super-resolved image. STED microscopy uses depletion of spontaneous fluorescence emission and beam shaping in confocal setting for surpassing the diffraction limit.

Super-resolution optical nanoscopy techniques (Fig. 3) overcomes the diffraction limit of light (~200 nm). These techniques allow imaging of nanoscale structure below 50 nm[45]. SIM, is a wide-field method with a typical resolution of around 100 nm, is one of the most promising super-resolution techniques due to its relatively high throughput and compatibility with common fluorophores[46,47]. As illustrated in Fig. 3, SIM projects a patterned illumination onto the sample and creates a Moiré pattern that allows high-frequency information to pass through the pass band of the collection lens. This method effectively doubles the resolution limit achievable by conventional microscopy. Advanced SIM techniques such as non-linear SIM have been also proposed to push this resolution limit further[48,49]. The superior resolution and faster acquisition speed of SIM enable its applicability in various applications including visualizing microvilli biogenesis ultrastructural details of placental tissue and intestinal tissue[10,50,51]. In these papers, SIM imaging is used to visualize nanometric structures in microvilli that can be used for diverse applications such as quantification of medical and junctional population[51]. It is important to note that SIM resolution can degrade with increasing sample thickness, such as in intestinal tissue. TIRF-SIM is an another advancement for achieving isotropic resolution enhancement but limited to only structures close to the glass surface [52].

Stimulated emission depletion (STED) microscopy is another super-resolution technique that achieve ultra-high resolution up to 10 nm[53]. STED is a point scanning technique that employs two lasers: an excitation beam and a doughnut-shaped depletion beam as shown in Fig. 3. The depletion beam suppresses the spontaneous fluorescence emission from the outer regions of the excitation spot and promote red-shifted stimulate emission which can be blocked using appropriate filters. This results in reducing the emission area and enhancing the final resolution. STED is a point-scanning confocal method; thus, speed is slow when scanning large area. STED has been used for various applications including the analysis of nanoscale organization of axonal proteins with myelin loops and microvilli[54]. In addition, STED has been used to study molecular organization of apical and lateral membrane domain to resolve the hexagonal packing of microvilli[55]. However, particular care should be taken to ensure proper fixation of membrane protein to achieve nanometric resolution and thus to visualize microvilli structures.

Finally, fluctuation-based super-resolution microscopy (FF-SRM)[56] is wide-field super-resolution technique that can be used for microvilli imaging. FF-SRM provides limited gain in super-resolution using 100-200 images. This technique analyzes temporal fluorescence intensity fluctuations caused by molecules blinking or transitioning between states. By correlating these temporal fluctuations, high-resolution images can be reconstructed to achieve resolutions ranging from 50 nm to 150 nm. Overall, each of these techniques has distinct advantages and limitations in biological imaging. For example, SIM offers relatively fast imaging which is suitable for many applications in live-cell studies however requires more complicated experimental set-up to generate accurate interference fringes and its manipulation. On the other hand, FF-SRM can be implemented using simple optical microscopy set-up, works with rather large number of fluorophores but does not provide very high gain in the resolution and can also suffer from reconstruction artefacts. STED provides ultra high-resolution and is effective for both fixed and live samples but often uses higher laser illumination as compared to other methods. These fluorescence-based techniques also require specialized sample preparation protocols. Below, we present the detailed steps of sample preparation and labeling protocols used in this study.

## 3. Sample preparations:

The fish used in this study are from the Tromsø Aquaculture Research Station, Kårvika, Tromsø, Norway. The fish was anesthetized using benzocaine (2.5ml in 10L water-ACD Pharmaceuticals AS). After 3-5mins carefully using a clean scalpel, an intestine segment was dissected into several pieces and stored in 8% PFA using PHEM marine solution (1,5X PHEM + 9% sucrose) and stored at 4°C until used for respective microscopic techniques.

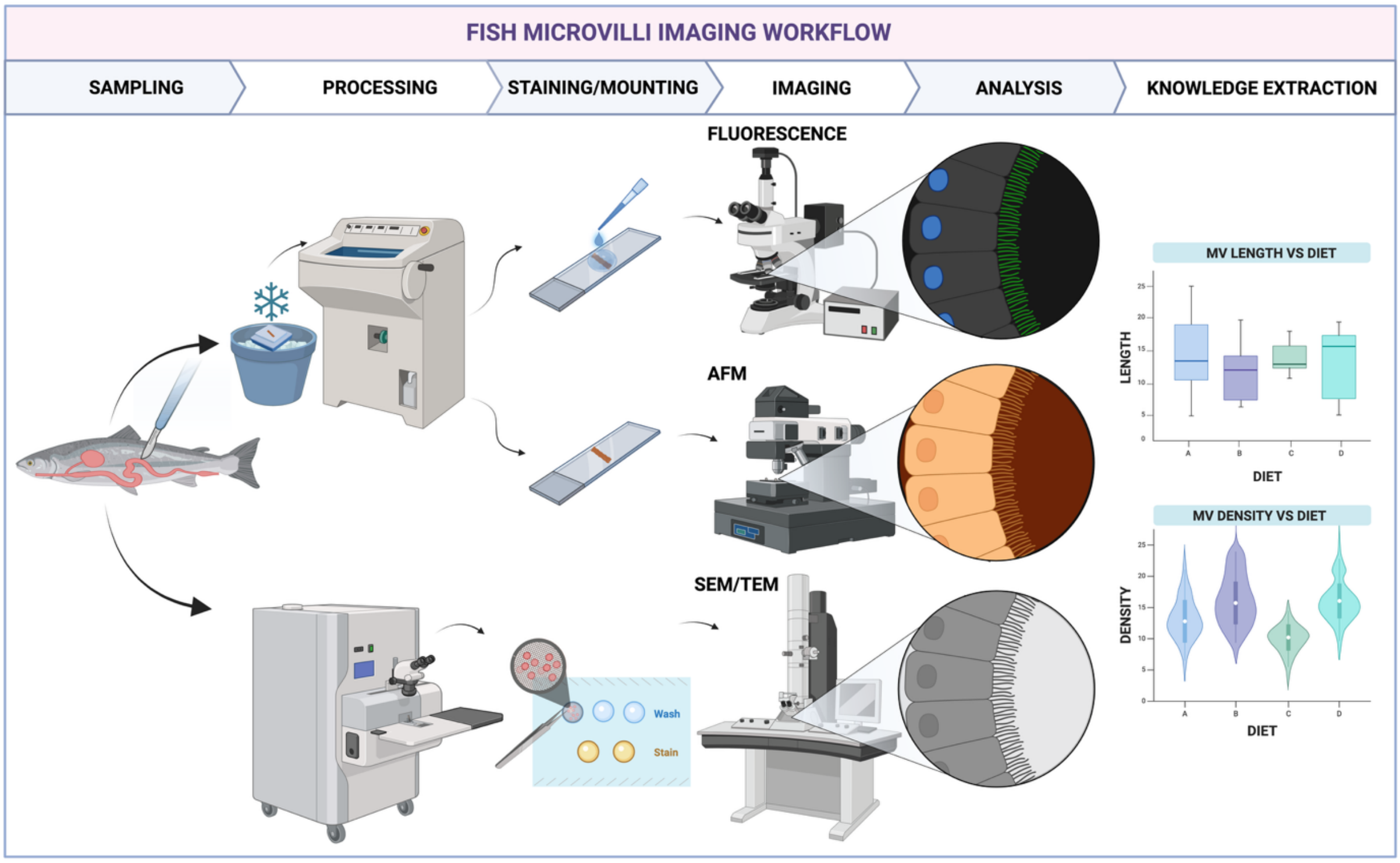

**Figure 4**: **Workflow illustration of the diverse preparation and imaging methods used in this study to visualize fish intestinal microvilli.** Upon anesthetizing, the fish intestine is dissected using a scalpel, washed, grossed, and subsequently preserved according to the protocol specified for each imaging technique. In this case, the Tokuyasu cryopreservation method was chosen for fluorescence-based microscopy and AFM, while standard electron microscopy (EM) preparation methods were used for both scanning electron microscopy (SEM) and transmission electron microscopy (TEM) imaging modalities. Although mainly focusing on the visualization of fish intestinal microvilli, this study paves the way for subsequent quantitative analyses of, for example schematically shown the impact of fish diet on the morphology and density of the intestinal brush border of farmed species such as Atlantic Salmon.

## Tokuyasu sectioning protocol:

Tissues were fixed in 8% PFA overnight and subsequently incubated in 0.12% glycine in PBS for 2 minutes to quench the free aldehyde groups. The tissues were infiltrated with 10-12% gelatin (FSG-0.5ml) for 1 hour in a rotator at 37°C. After the incubation, the tissues were transferred to 2.3M sucrose in water and left overnight in a rotator at 4°C. Thereafter, the samples were placed on cryosticks and stored inside a cryotube in liquid nitrogen. For sectioning, the samples were taken from the nitrogen tank and sliced using a cryo ultramicrotome at 400 nm thickness. In this study, BF, QPM, chip based TIRF, confocal, SIM, STED, and AFM images are acquired using Tokuyasu prepared samples.

## SEM microscopy:

For SEM, briefly, the samples were fixed overnight in 2,5% glutaraldehyde in PHEM buffer in phosphate buffer, (pH 7.2). After washed with 0.15 M PHEM marine buffer, the coverslips containing the tissues were treated with 1% Osmium tetroxide in doubled distilled water(ddH2O) for 1.5 hour and dehydrated in series of ethanol (30%, 60%, 90% for10 min each, 5 times 100% ethanol for 10 minutes), and after dried in a critical point drier (LeicaEMCPD300), mounted and sputter-coated with 10 nm gold/palladium alloys. Specimens were examined using a commercial SEM (Gemini, Zeiss), run at 2.00 kV, and imaged at different magnifications.

## TEM microscopy:

For TEM, tissues were fixed for 4 hours in 4% PFA buffer and washed with PHEM marine 2 times at 15-minute intervals. Then the tissue was incubated with 1% OsO4 in dd H20 for 90 minutes and repeat the 15 minutes washing with PHEM marine buffer and a quick rinse with ddH2O. The tissue was incubated with 2% uranyl acetate for 90 minutes and followed by dehydration in a graded series of ethanol (30%,60%, 90%, 96%, 2 × 100%). The tissue was treated with acetone as an intermediate step before infiltration with an Epon substitute (AGAR 100 resin, Agar Scientific, Stansted, England) and polymerized at 60 °C overnight. Ultrathin sections (70 nm) were made using a Leica Ultracut S Ultramicrotome (Vienna, Austria) with a Diatome diamond knife (Biel, Switzerland). The sections were mounted on carbon-coated formvar films on copper grids and contrasted with 5% uranyl acetate for 8 minutes and Reynolds lead citrate for 5 minutes. Micrographs were taken on a Jeol 1010 JSM (Tokyo, Japan) with a Morada 11 Mpixels digital camera (Olympus).

## AFM microscopy:

Samples were measured in PBS at 25ºC using PetriDish heater™ (JPK Instruments/Bruker). MSCT (Bruker) cantilevers (k = 0.03 N/m, nominal tip apex radius of 10 nm) were used for the imaging in Quantitative Imaging (QI) mode, according to the methodology described before for LSEC fenestrations[57,58]. Briefly, each image was acquired by performing multiple force curves in each pixel(px)/point of the image that were translated into the images of topography and stiffness, where stiffness served only as a high contrasted image allowing detection of glass substrate (stiff – bright) and cell (soft – dark). Load force was in the range of 600-800 pN. The length of the force curves (the z range) and the acquisition speed were in the range of 2.0-2.5 µm and 140-160 µm/s, respectively. Before measurements, the AFM detector sensitivity and spring constant was evaluated using non-contact method. Specimens were examined using a commercial AFM (Nanowizard IV, JPK Instruments/Bruker).

## Bright field and quantitative phase microscopy:

A homebuilt system was used to acquire both bright field and QPM images of the microvilli, as previously described in the methodology section for the QPM system[17,18]. Briefly, a partially spatially coherent Linnik-type QPM setup was employed to capture off-axis interferometric images. The bright field images of the tissue sample were acquired by blocking the reference arm of the Linnik system. A 660 nm laser and a 60X, 1.2 NA objective lens (Olympus) were used for imaging in both modes. The samples were prepared on top of a silica wafer to obtain high-contrast interferometric fringes. PZT is used in the reference arm to acquire five phase shifted images of the sample. These images were reconstructed by using five-phase shifting algorithm[59].

## Structured illumination microscopy:

Commercial super-resolving SIM (DeltaVision/OMXv4.0 BLAZE, GE Healthcare) with a 60X, 1.42 NA oil immersion objective lens (Olympus) is used for microvilli imaging. Raw datasets acquired from the SIM system were computationally reconstructed using SoftWoRx software (GE Healthcare). Further details about the SIM system can be found here[60].

## Chip-based TIRF/Chip-based FF-SRM:

A homebuilt photonic chip-based system is used to acquire TIRF and FF-SRM images of microvilli. Details of the system is described in our previous paper[41]. Briefly, chip TIRF system was assembled using upright microscope (BXFM, Olympus) and a photonic chip. The sample was placed on top of the photonic chip and fluorescence emission of the samples was achieved via evanescent field as described previously[44]. The fluorescence signal was captured through a simple upright microscope. In our study, the illumination wavelength used were 640 and 561nm. A 60X 1.2NA objective lens is used to collect the raw images with an exposure time between 10-50 msec. Further details about photonic chip, microscope, alignment, data acquisition, and post processing can be found here[41,44,61,62].

## Confocal and STED:

Confocal and STED images of the microvilli samples were acquired using the STEDCON system. This system can capture multicolor confocal and STED images and is equipped with five different lasers: 405 nm, 488 nm, 561 nm, 640 nm, and 775 nm. In this study, we use 60×, 1.42 NA objective lens to acquire the images. More details about the system can be found here[63].

# 4. RESULTS AND DISCUSSIONS

## 4.1. Diffraction limited fluorescence-based techniques for microvilli imaging

Fluorescence based optical imaging techniques were used to investigate the structure of fish intestinal microvilli. First, diffraction-limited techniques such as wide-field fluorescence microscopy, TIRF microscopy, and confocal microscopy were utilized for imaging as they provided high-throughput and easy to implement, albeit with a diffraction limited resolution as outlined in Fig. 1. These methods provide a maximum spatial resolution of 200-250 nm. Figure 5(a) shows the wide-field fluorescence image of fish microvilli acquired by deconvolution microscopy system. In this image, the actin region of the microvilli is highlighted using phalloidin ATTO 647N. Imaging actin is interesting as it forms the structural core of microvilli that plays a significant role in providing mechanical support and stability. Imaging the actin allows for the quantification of microvilli physiology and the analysis of their remodeling under different conditions[13]. This approach is particularly useful for understanding structural changes associated with fish nutrition, development, and onset of diseases.

Previous studies have successfully employed fluorescence-based techniques to investigate the structure and function of intestinal M cells[13]. Another study demonstrates the applicability of imaging tubulin and actin to identify cilia and microvilli[64]. An interesting application of fluorescence microscopy is to detect the effect of microplastics in fish organs. Confocal microscopy is applied for this purpose to demonstrate the uptake and translocation of microbead plastic particles[65]. On-chip TIRF and confocal images of the microvilli are shown in Fig. 5 (b, c). Their zoomed regions of interest are shown in Fig. 5($b_1$, $c_1$) to show the capacity of these techniques to resolve the microvilli structures. Since the width of microvilli ranges between 50–200 nm, these techniques can barely resolve them due to their inherent diffraction-limited nature. Nonetheless, the zoomed image reveals that confocal imaging offers slightly better resolution compared to on-chip TIRF and widefield fluorescence imaging. It is expected as the confocal technique employs a point scanning approach to remove out-of-focus light to achieve the best possible resolution among diffraction limited techniques.

In addition, on-chip TIRF experiments were performed with water immersion objective lens 60x 1.2 N.A. and the confocal and deconvolutions experiments were conducted using 60x 1.42 oil immersion objective lens. Thus, a small offset in the resolution can be seen due to the difference in the N.A. However, due to point scanning, confocal imaging compromises temporal resolution when imaging large sample. On the other hand, on-chip TRIF is useful where wide field, optical sectioning is needed to quantify fine features in thin tissue sections, such as intestine. On-chip TIRF is a powerful optical technique that provides high-contrast signal over a large field of view while suppressing unwanted signals from the out-of-focus layers of the sample. In addition, it is not a point scanning method therefore, the temporal resolution is comparatively higher than the confocal imaging. One notable advantage of using these imaging techniques is their ability to capture morphological changes across a larger field of view[66]. Overall, the diffraction limited microscopy methods are suitable for wide field imaging and to quantify overall morphology but limited to quantify nanometric features of microvilli.

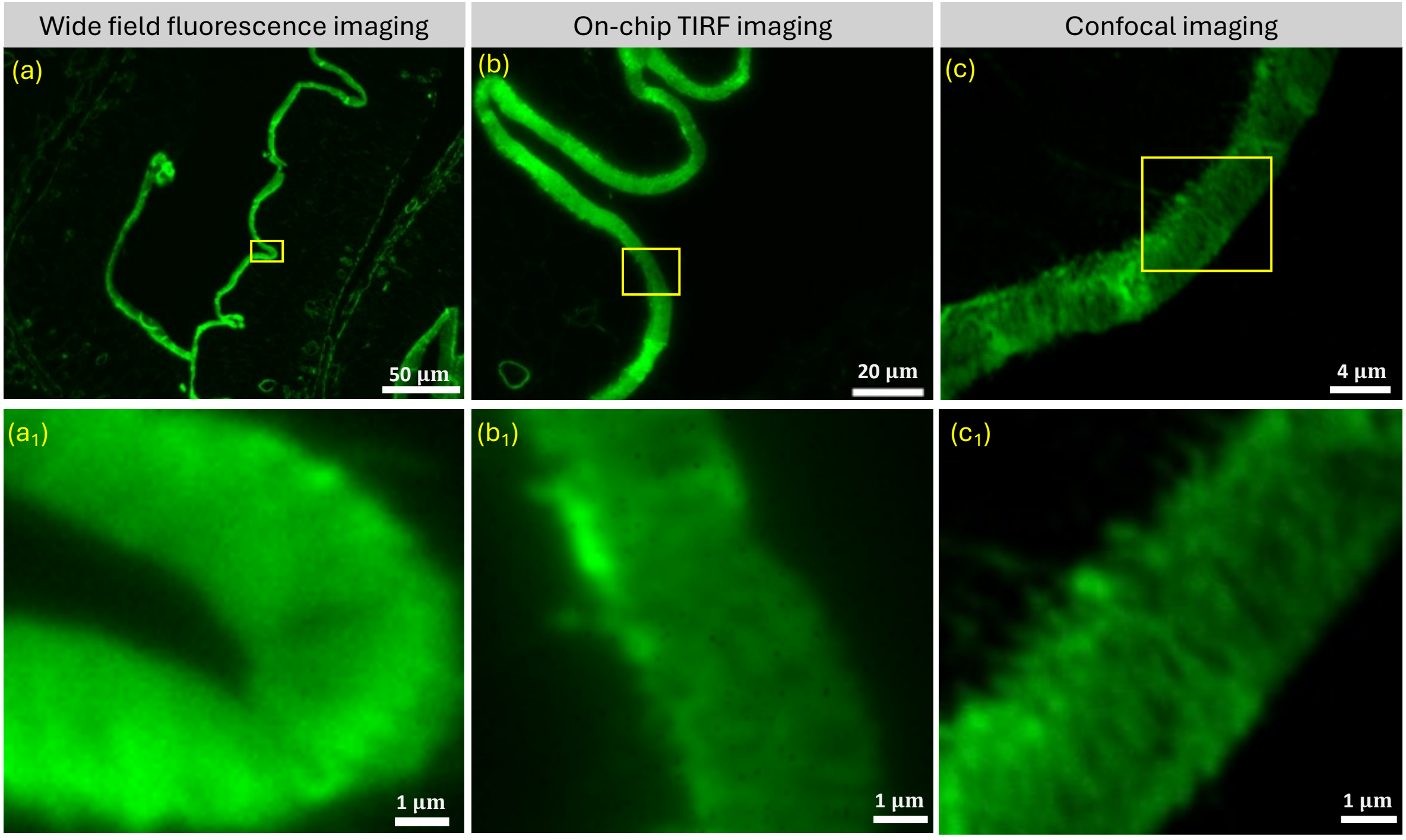

**Figure 5**: **Diffraction limited fluorescence images of fish microvilli**. (a) wide field fluorescence imaging, (b) on-chip TIRF imaging, and (c) confocal imaging technique. The zoomed-in field of view from each technique is shown to compare their capabilities in imaging microvilli. Note that wide-field imaging provides a large field of view but with limited resolution. In contrast, confocal imaging offers improved resolution at the expense of a smaller field of view compared to their counterparts. These techniques can achieve a best possible resolution of approximately 200–250 nm.

## 4.2.Super-resolution fluorescence nanoscopy techniques for microvilli imaging

In contrast to the diffraction limited techniques, super-resolution optical microscopy, also commonly referred as optical nanoscopy managed to advance the experimental systems and computational algorithms to push the resolution limit down to sub-20 nm. Here, we show the utility of super-resolution optical microscopy techniques in imaging microvilli structures as shown in Fig. 6. We used three different techniques, STED, SIM, and chip-based SR-FFM. These techniques are also compared in terms of imaging throughput and resolution.

STED provides the best possible resolution among existing optical nanoscopy; therefore, it can be an asset for imaging microvilli as illustrated in Fig. 6 (a). The zoomed view of the STED image is shown in Fig. 6 ($a_1$) where length and distribution of microvilli is clearly visible. We also plotted the line profile along the yellow line to quantify the width of microvilli using STED. The line profile shows that STED can clearly resolve the spatial distribution of individual microvilli. The width of each microvillus in the STED image is found to be between 60 and 140 nm. In the past, STED has been successfully used for imaging microvilli[11,12,67]. In principle, STED can provide very high spatial resolution, however, this gain in resolution comes at the cost of phototoxicity, low temporal resolution and low throughput when large areas must be scanned.

The sweet spot between resolution and throughput in optical nanoscopy can be achieved by using SIM. The SIM method achieves close to 100 nm resolution and offers high-speed imaging making it suitable for both fixed and live cell imaging. The SIM image of microvilli and zoomed field of view are shown in Fig. 6(b, $b_1$). It is evident from the zoomed image and the line profile that the resolution of SIM is slightly lower than that of STED. Note that the best achievable resolution with linear SIM is approximately twice the diffraction limit, i.e., around 100 nm, whereas STED can achieve resolutions down to 20 nm. Nevertheless, it is still possible to quantify several microvilli in SIM with feature sizes larger than 100 nm. A recent study has demonstrated the periodic organization of microvilli in both 2D and 3D using interreference based SIM technique[12]. Other studies have shown using confocal and exploiting the potential of SIM for visualizing microvilli[10,50]. In addition, experimental and computational power of SIM is also shown and compared with electron microscope by imaging ultrafine features of microvilli[68].

Next, we explored photonic-chip based SR-FFM which is a comparatively new technique that supports high-resolution, high throughput, and multimodal imaging[41]. The multi-mode illumination using a photonic chip generates the sparsity in the emission light that can be exploited to generate super-resolution images using SR-FFM. Here, we used SR-FFM to generate super-resolution images of intestinal microvilli using chip based nanoscopy (Fig. 6(c, $c_1$)). The zoomed image in Fig. 6($c_1$) shows the microvilli structures. Although, it can help to improve the contrast of an image compared to the diffraction limited technique, the fluorescence fluctuation-based techniques can barely help to resolve microvilli structures as can be seen in the line profile. One possible reason could be the densely packed structure of microvilli, where fluorophores are located very close to each other. The overlap of fluorophore signals makes it difficult to distinguish individual intensity fluctuations, which makes it hard to achieve super-resolution. Nonetheless, it is a well-known technique for achieving sub-diffraction resolution through low excitation intensities, fast acquisition, and high throughput imaging.

In addition, the chip-based techniques reduce the complexity of the traditional nanoscopy system and thus provide a simplified solution for imaging nanometric structures such as microvilli over larger field of view. However, the trade-off for this extended field of view in chip-based nanoscopy is a lower resolution compared to SIM and STED. SIM achieves a resolution of approximately 100 nm that might be sufficient for visualizing microvilli structures whereas STED offers the highest resolution (~20 nm) among optical microscopy techniques. The next section covers the comparison of conventional and cutting-edge label-free imaging technique for microvilli imaging.

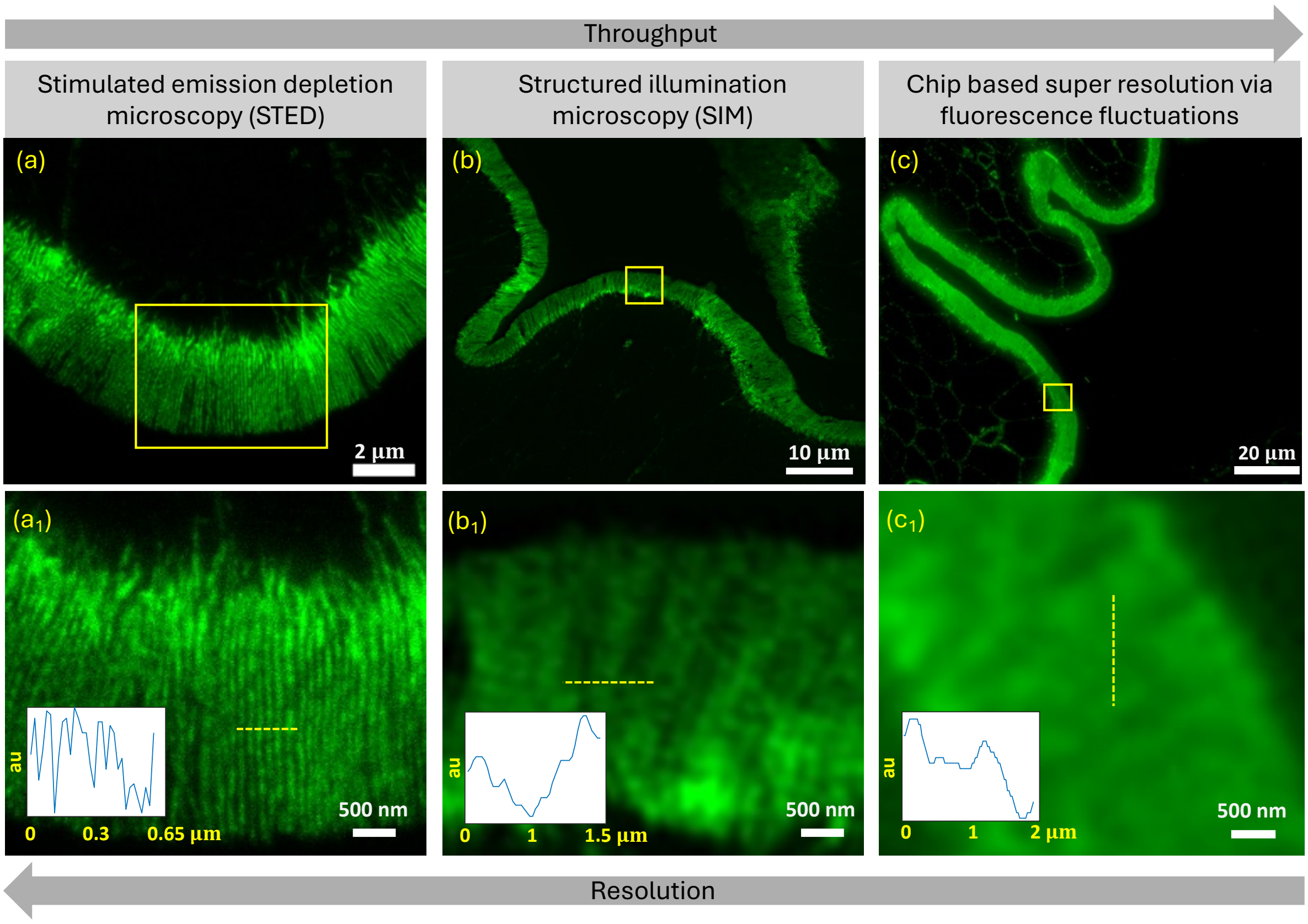

**Figure 6**: **Super-resolution fluorescence imaging of fish microvilli**. (a) Stimulated emission depletion microscopy (STED), (b) structured illumination microscopy, and (c) chip based super resolution via fluorescence fluctuation. Chip-based techniques reduce the complexity of the traditional nanoscopy system and provide a simplified solution for imaging microvilli whereas STED offers the highest resolution (~10 nm) among optical microscopy techniques. ($a_1$-$c_1$) The zoomed field of view highlights the capability of fluorescence nanoscopy techniques to resolve microvilli structures. The line profile along the yellow line clearly shows that among all these techniques, STED most clearly resolves distinct microvilli features. The x-axis of the line profile represents spatial distance in microns, and the y-axis shows the normalized intensity.

## 4.3. Label-free quantitative techniques for morphological imaging of microvilli

The label-free imaging methods provide sample information in their natural state without using exogenous labels. Ease of sample preparations in label-free imaging makes it compatible for diverse applications. Here, we used two different label-free techniques namely bright field microscopy and QPM. The bright field image of the fish microvilli sample is shown in Fig. 7(a). Bright field is a conventional and most versatile label-free method of imaging biological samples. It is a qualitative method to visualize micrometer-scale features in the sample as can be seen in Fig. 7(a). However, it cannot quantitatively differentiate between the subcellular features presented across the sample.

QPM has a potential to quantify these subcellular features as shown in Fig. 7(b). The similar FOV is imaged under QPM to see the complementarity of these two techniques. The color map in the QPM represents the phase map of the sample in radians. The phase map is a combined information of the refractive index and the thickness of the sample. The yellow arrow in the QPM image indicates the optical thickness of the microvilli region, which quantitatively shows that the microvilli region in submicron thick. It is important to note that we used a highly spatially sensitive QPM system which can detect sub-100 nm changes in the axial direction. In other words, highly sensitive QPM techniques can quantify the refractive index map of the microvilli with nanoscale axial sensitivity. Here the sensitivity refers to a change in the optical path length in the specimen can be picked up with nanoscale accuracy.

The nanoscale sensitivity of QPM along the axial direction can be useful to differentiate between distinct types of microvilli structures. Also, it is important to note that QPM is still a diffraction-limited technique and lacks chemical specificity, which limits its ability to quantify nanostructure of the microvilli in the spatial direction. Nonetheless, the complementarity of QPM can be useful in correlative imaging, where QPM can quantify nanoscale thickness of the microvilli area and super-resolution optical imaging techniques such as SMLM/SIM can be useful to provide super-resolution images of the microvilli along the spatial direction[18].

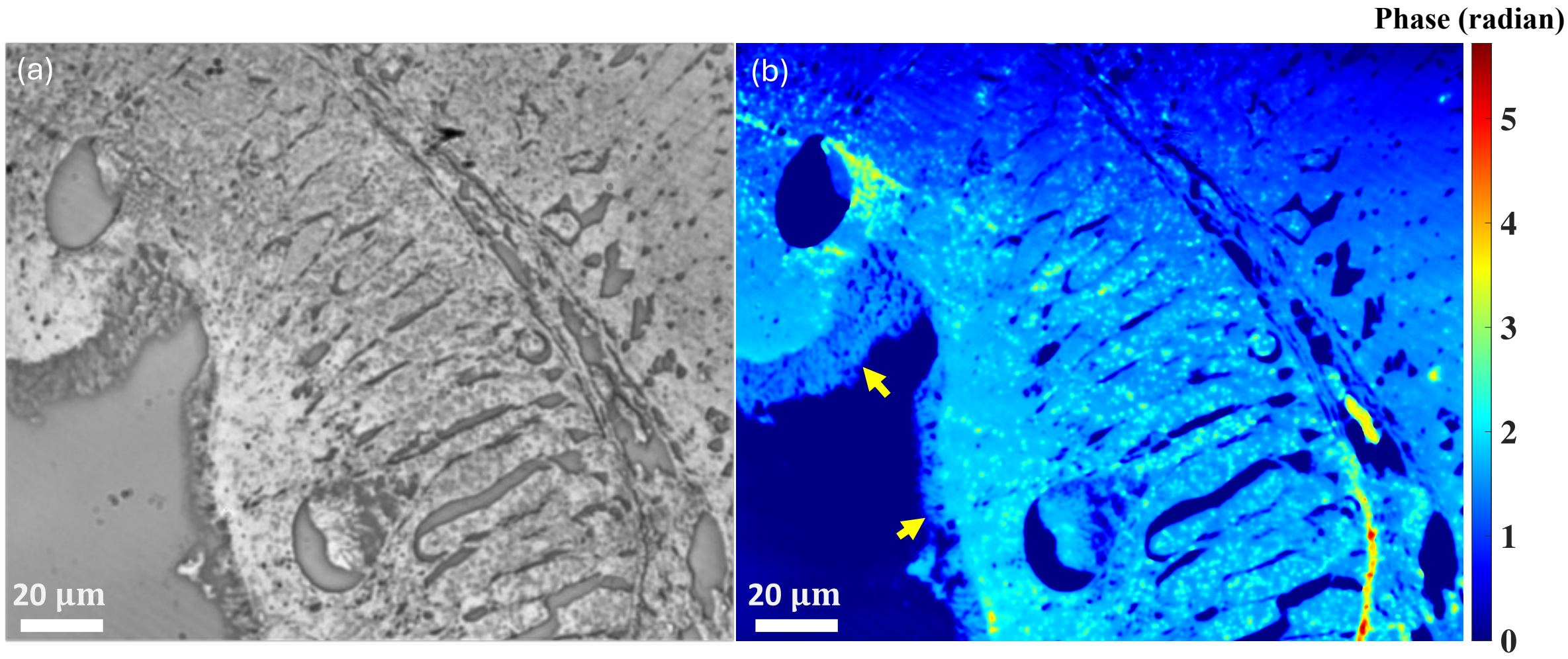

**Figure 7**: **Label-free imaging methods for morphological imaging of microvilli**. (a) Bright field image of fish microvilli to extract qualitative information. (b) Quantitative phase imaging (QPI) of microvilli that represent the combined information of refractive index and thickness of the sample. Since, it is a diffraction limited technique, it cannot distinguish between nanometric microvilli structures. Interestingly, the nanoscale sensitivity of QPI retrieve of nanometric thickness in microvilli structures, as highlighted by the yellow arrow. The color map in the QPI image represents the phase map in radian.

## 4.4. SEM, TEM, and AFM methods:

Next, we compared the electron and the force microscopy methods that can provide the best resolution. High resolution capabilities of these methods allow imaging of nanometric structures of microvilli. Figure 8 (a) shows SEM image to visualize the surface morphology of microvilli. The SEM image reveals the 3D ultrastructure and densely packed arrangement which is particularly advantageous for studying surface topography with 1-20 nm resolution. SEM enables examination how microvilli length and density vary depending on dietary habits (e.g., carnivorous vs. herbivorous diets in animals) or adaptive responses (e.g., in fish, microvilli may change in response to salinity levels or environmental stress) [69,70]. SEM is traditionally the preferred method to visualize the microvilli. SEM generates 2D images that can convey a sense of depth and three-dimensional structure and integrity, while TEM is used to e.g., assess the length and widths [69,70]. On the downside, the sample preparations steps for SEM imaging are quite tedious as it required coating the sample with a conductive layer to enable the interaction of electrons with the sample surface. In addition, SEM is limited to surface imaging, required fixation, dehydration, and sputter-coating which may alter the biological structures. Finally, SEM provides a topographical view and therefore cannot be used to accurately measure the thickness of microvilli or the gaps between them.

TEM (Fig. 8 (b)) provides superior resolution as compared to SEM enabling better visualization of microvilli ultrastructure. For TEM imaging, sample must be ultrathin-sectioned (~70 nm) after being embedded in resin, followed by staining with heavy metals like osmium tetroxide to enhance electron contrast. TEM provides high-resolution images of microvilli's length, density, and distribution, which can vary depending on the cell type and its specific function and adaptation. The line profile along the yellow line in TEM image show that it reveal the precise arrangement of these actin filaments within microvilli, which is crucial for understanding their structural stability and function. TEM can also provide detailed images of the lipid bilayer and membrane-associated proteins of the microvilli, giving insight into their functional properties, such as ion transport, receptor signaling, and nutrient absorption[71]. Therefore, SEM and TEM are considered as a conventional method for studying nanostructure features in microvilli.

Next, we used AFM, which can deliver similar resolution as SEM and TEM but does not require complex sample preparation or conductive coating that makes it suitable for biological imaging. It provides a topographical map of the microvilli with a color-coded height scale as shown in Fig. 8 (c). AFM has been used previously to measure the topography of fish scales and fish skin epithelial cell elasticity dynamics[72,73], but not yet on intestinal microvilli. The resolution of AFM lies in between electron microscopes and optical microscopes. Based on stiffness contrast data, it is possible to provide quantification of microvilli using AFM. Moreover, stiffness/elasticity (Young's modulus) can be an additional parameter that might be valuable for detecting differences at the nanoscale. While it offers an advantage of live cell imaging and nanoscale height variation, AFM's imaging speed is relatively slower, and it is therefore less effective for high throughput imaging. Overall, SEM, TEM, and AFM are well suited for high-resolution imaging and to measure the mechanical properties of the objects. On the downside, AFM, SEM and TEM lacks quantitative imaging, chemical specific imaging, and high throughput imaging that can be complemented by fluorescence based optical microscopy methods.

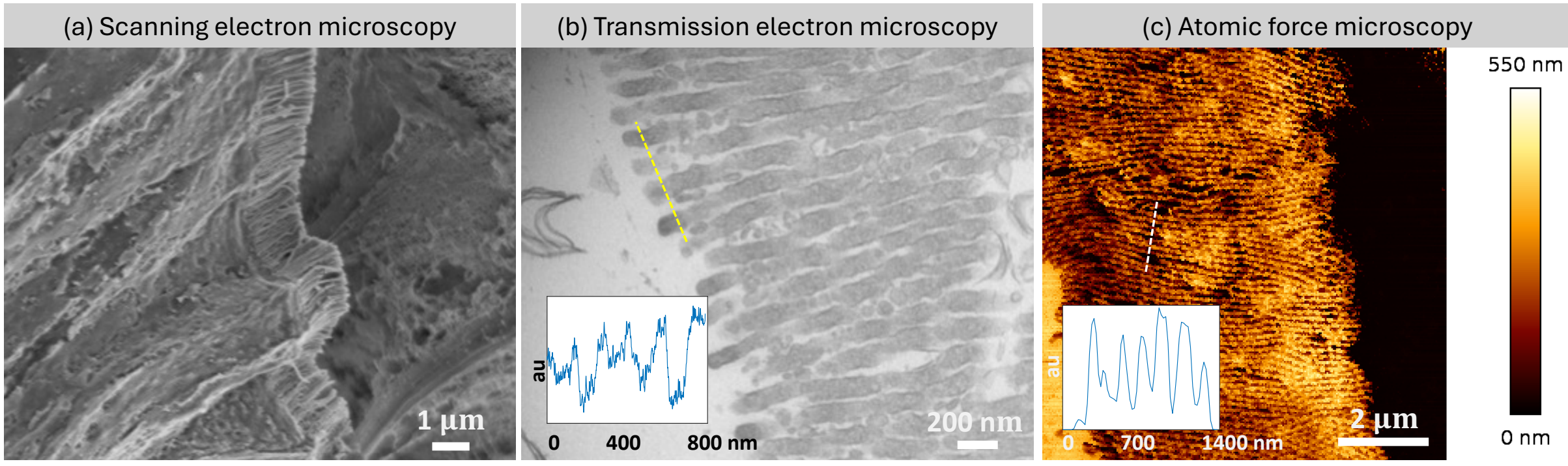

**Figure 8**: **Super-resolved images of microvilli** (a) scanning electron microscopy, (b) transmission electron microscopy, and (c) atomic force microscopy. The yellow line along the TEM image and white line along the AFM image shows the thickness of individual microvilli. SEM provides a topographical image and therefore cannot be used to measure microvilli thickness. The width of microvilli is found to be 50-150 nm.

## 5. Conclusion

This article summarizes an overview of key imaging techniques encompassing, optical, electron, and atomic force microscopy for studying epithelial tissues in fish. It highlights distinct aspects of these imaging systems (as shown in Fig. 1) that play a vital role in understanding the structural and functional properties of microvilli. Improvement in these parameters such as pushing the resolution limit, ease of sample preparation, and temporal resolution may further advance their capabilities for quantifying microvilli or any other biological applications. As different methods provide different distinct advantages adopting multimodal approaches could be a valuable option. Similarly, understanding the two-way complementarity benefits between labeled and label-free optical imaging systems will open new opportunities. Such as multi-modal QPM with fluorescence assisted nanoscopy will allow overall context via refractive index tomography of the tissue sections and nanoscopy can be used to see delicate details of the micro-villi providing holistic understanding of the sample. Both labeled and label-free systems offer unique advantages and provide distinct information. On way to bridge the gap is to employ artificial intelligence to create a unified software platform that leverages the strengths of both systems. However, a careful examination must be required to ensure its effectiveness and reproducibility of artificially generated complementary information. Despite these advancements both in label-free and fluorescence-based microscopy, both approaches are not even close to electron microscopy and AFM in resolving nanometric features of cells and tissues. Another approach worth investigating in the future would be expansion microscopy which has emerged as an powerful approach that can utilize the conventional microscopy methods to provide super-resolution by physically and isotopically expanding the biological sample.

## Acknowledgments:

Authors acknowledge support and funding from UiT—The Arctic University of Norway and the Research Council of Norway (grant no. 325159 and 352764). The authors would like to acknowledge Randi Olsen and Kenneth Bowitz Larsen at the Advanced Microscopy Core Facility at UiT – The Arctic University of Norway for their assistance is sample preparation for EM and STED microscopy. We would like to acknowledge Sebastian Acuna for providing inputs with fluorescence fluctuation with super resolution techniques.

## Conflict of interest

BSA has applied for a patent for chip-based optical nanoscopy, and he is co-founder of the company Chip NanoImaging AS, which commercializes on-chip super-resolution microscopy systems. Other authors declare no conflicts of interest.

## Author contribution

BSA and RAD conceived the idea and supervised the work. LEVH and DVT mainly prepare the samples, coordinated with the experiments. AB, LEVH, DVT acquired and analyze the images. BZ and LEVH acquired the AFM images. AB mainly wrote the manuscript, and all authors contributed to revising different sections.